# A Partition-Driven Integrated Security Architecture for Cyber-Physical Systems

Yahya Javed*, Muhamad Felemban*, Tawfeeq Shawly*, Jason Kobes† and Arif Ghafoor*

*School of Electrical and Computer Engineering, Purdue University, West Lafayette, IN
†Northrop Grumman Corporation, Washington, D.C.

*Emerging cyber-physical systems incorporate systems of systems that have functional interdependencies. With the increase in complexity of the cyber-physical systems, the attack surface also expands, making cyber-physical systems more vulnerable to cyber-attacks. The functional interdependencies exacerbate the security risk as a cyber-attack that compromises one constituent system of a cyber-physical system can disseminate to others. This can result in a cascade effect that can impair the operability of the whole cyber-physical system. In this article, we present a novel security architecture that localizes the cyber-attack in a timely manner, and simultaneously recovers the affected cyber-physical system functionality. We have evaluated the performance of the architecture for advanced metering infrastructure-based pricing cyber-attacks scenario. The simulation results exhibit the effectiveness of the proposed architecture in containing the attack in terms of system availability and its impact on the electric load distribution in the power grid.*

The increase in popularity of automated systems has made cyber-physical systems (CPSs) find their application in many diverse domains, including electrical power grid, automotive systems, smarthome systems and healthcare systems. Emerging CPSs integrate heterogeneous systems that have functional interdependencies. These integrated heterogeneous systems effectuate discrete functions and collaborate in an intricate fashion to perform the overall functionality of the CPSs.[1] Modern CPSs are connected to the internet and employ cloud-based infrastructure for their computational requirements. Moreover, the systems of a CPS systematically communicate with each other and the field devices of a CPS are commonly provided with one or more interfaces for maintenance purposes.[2] Nowadays, CPSs are capable of performing multiple diverse functions as compared to the legacy systems. However, such capability can result in an increased system complexity. From the systems engineering perspective, the increase in system complexity makes it more challenging to provide service guarantees. Similarly, from the security perspective, added complexity widens the attack surface.[3] The functional interdependencies among the systems of CPS further exacerbates the security risk. An attack that is initiated at one sub-system of a CPS can disseminate to others. This can result in a cascade effect of compromising one system after other and eventually impairing the functionality of whole CPS in a short amount of time.

Some CPS security frameworks mitigate the security risk by promptly patching the system components for newly discovered vulnerabilities, in addition to implementation of strict authentication protocols for communication with the CPS.[4] This solution based on the target hardening concept is proving to be less effective as cyber-attack incidents reported over the years indicate a continuous increase in their sophistication. Another solution is based on the intrusion prevention approach, in which the security framework employs an intrusion detection system (IDS) that generates alerts when it identifies an anomaly in the working of CPS components. On the other hand, intrusion prevention systems (IPSs), in addition to detection, can deploy some countermeasures, such as disconnecting the attacker's communication with the CPS, and halting the operation of affected CPS components.[5]

The performance of IPSs is tightly coupled with the IDS efficiency. IDS is usually evaluated in terms of false positive rate (FPR), false negative rate (FNR) and detection delay time. High FPR indicates significant number of false alarms that can make the behavior of the IPS unpredictable while high FNR implies an ineffective IPS. Detection delay time is often meted with less importance in evaluating the performance of an IDS, but it has a significant effect on the operations performed by the IPS. Considering a scenario in which the intent of the attacker is to compromise as many components of the CPS as possible. The attacker orchestrate the attack by systematically compromising one component after the other. The average time taken by the attacker to compromise a component and move to compromise the next one corresponds to the attack propagation speed. On the other hand, the average time taken by the IDS to identify if a component has been compromised corresponds to the detection speed. If the attack propagation speed is more than the detection speed then the attacker can continue to compromise the CPS components while the IPS would try to catch-up until all of the target components are compromised. This "catch-up" process is evident when the strategy of the attacker is to outpace the response or defense mechanism of the system rather than evade detection.

The contribution of our work is multifold. First, we present an integrated security architecture that combines the concepts of intrusion prevention and recovery to manage cyber-attacks while maintaining high operational availability of CPS. Second, we propose a novel multi-level partitioning mechanism that partitions the CPS components based on their functionality and damage containment requirements. The partitions help to contain the damage by localizing the cyber-attack in the region of its inception. The response mechanism of the architecture effectively utilizes the partitions to counter attacks involving high attack propagation speeds as well. Finally, we have simulated advanced metering infrastructure (AMI)-based pricing cyber-attacks scenario, and demonstrate the performance of the proposed architecture in terms of operational availability ($\omega$), damage extent ($\delta$) and average energy load (AEL) consumption.

## An Integrated CPS Security Architecture

The proposed security architecture and its interaction with a general CPS framework is depicted in Figure 1. The architecture has four constituent systems: partition manager (PAM), intrusion response system (IRES), intrusion recovery system (IREC) and performance monitor (PEM). The architecture utilizes the service of an intrusion detection system (IDS) that generates alerts upon detection of a cyber-attack. The architecture also uses the operational logs of CPS and assumes that they are updated in realtime with CPS component activities such as data flow between components with timestamp. We discuss the constituent systems of the architecture and their internal modules in detail starting with a discussion on the intrusion detection process.

**Intrusion Detection System**
Several intrusion detection techniques corresponding to different threat models exist in literature. IDSs employ diverse methods to identify the anomalies in the operations of the systems they are designed to protect. An IDS performs two major operations: collection of data and analysis of the data to detect malicious activities. CPS centric IDSs collect data by monitoring component activities or by observing CPS network traffic. The detection techniques are either signature-based or behavior-based. In signature-based approach the IDS looks for malicious pattern to make detections. In behavior-based approach the IDS compares every new observation with normal system behavior it has learnt during the data collection phase.[6] Some IDSs also provide limited IPS capabilities e.g. snort[7] combined with firewall configuration feature can filter any malicious stream from a link. This technique of filtering out malicious traffic refers to as negative filtering. It is also possible to do positive filtering in which only traffic meeting certain criteria is allowed. Firewalls are usually deployed on a central communication link to ensure only benign traffic flows through the link and it is not feasible to deploy them on every communication link of the system. For our proposed architecture, we assume that an IDS is available and works across all layers of the CPS. We assume that the IDS can report the cyber-attack type, start time and detection time of cyber-attack, and compromised components of CPS. We also assume that the firewalls can be deployed on certain communication links of the CPS.

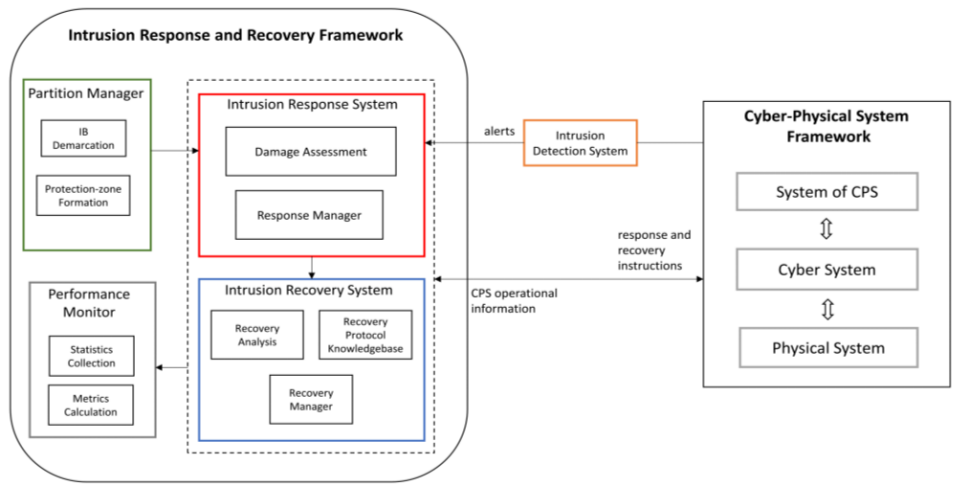

Figure 1. Proposed CPS security architecture and its operations across all CPS layers.

**Partition Manager**
PAM creates two levels of partitions for the CPS. The first level of partitioning involves demarcation of intrusion boundaries (IBs). The objective of the IBs is to contain damage propagation across IBs. The second level of partitioning involves the formation of protection-zones. The objective of protection-zones is that within an IB, components affected by the attack needs to be isolated quickly so that the impact of the attack is confined to affected components only.

**Intrusion Boundary Demarcation**

The CPS components that collaborate to provide a specific CPS function are included in the same IB. It is possible that a CPS component handles multiple functions while the same CPS component is shared among multiples IBs. An IB can encompass more than one functions as well. To create IBs, CPS topological structure and CPS component to functionality mapping is used, and are assumed to be available to the architecture. Figure 2(a) shows advanced metering infrastructure (AMI) architecture and Figure 2(b) shows the division of AMI components in different IBs. The smart meters that are connected to the same data concentrator perform meter_ops_nan function, and form a neighborhood area network (NAN). Therefore, they are part of the same IB. Similarly, the AMI components at the enterprise system layer such as meter data management system (MDMS), consumer information system (CIS), outage management system (OMS), distribution management system (DMS) and geographic information system (GIS) are mapped to global_meter_management function, and can be placed in the same IB. The idea behind IBs is that an attack inside an IB is fully contained in it by design. The communication among different IBs is carried out by dedicated links or in some cases through shared components. The inter-IB communication can be ensured to be not malicious by applying a positive firewall filter policy on the inter-IB communication links or on the outgoing traffic from the shared component. To minimize the firewall deployment overhead, the number of inter-IB communication links should be minimum. To find an optimal IB configuration, an optimization problem similar to classical graph partitioning problem[8] can be formulated that minimizes the edge cut among IBs.

**Protection-zone Formation**

IBs can contain thousands of CPS components and an attack starting from a single component in an IB can potentially compromise all of the components in that IB due to functional dependencies. Figure 2(c) shows an example AMI topology enclosed in an IB. In order to contain the damage within an IB, we introduce a second level of partitions known as protection-zones. The PAM groups CPS components of an IB in almost fixed-size protection-zones such that the chances of inter protection-zone damage propagation are minimized. The response system of the architecture utilizes the protection-zone information to effectively localize the damage in an IB. The protection-zone creation process is formulated as a single-objective multiple constraints optimization problem with the goal to minimize the inter protection-zone communication.

The CPS components in an IB and the interconnections among them can be modeled as an unweighted non-directional graph $P$, as shown in Figure 2(c). Let $P = \{V, E\}$ where $V$ is the set of vertices modeling CPS components and $E$ is the set of edges modeling the communication links between CPS components. Let $P_i$ be the ith subgraph, such that $P = \cup_{i=1}^{k} P_i$ and $P_i \cap P_j = \emptyset$. For each $v \in P_i$, let $Q_i = \{P_j \mid \exists\, e(v,u)\ and\ u \in P_j\ where\ j \neq 1\}$ i.e. $Q_i$ is the set of neighboring subgraphs that contain a vertex $u$ that is adjacent to $v \in P_i$ whereas there exist an edge $e(v,u)$ between $u$ and $v$. The problem is to partition the graph $P$ into $k$ balanced subgraphs with minimum edge cut such that a vertex at the boundary of a subgraph is at most connected $1 + \eta$ other subgraphs. Let $\otimes_{ij} = \{e(v_i, v_j) \mid v_i \in P_l\ and\ v_j \in P_k, \forall k \neq l\}$, i.e. $\otimes_{ij}$ is the set of edges extending from one subgraph to another. The optimization problem is as follows.

$$\min \sum_{i \neq j} |\otimes_{ij}| \qquad (1)$$

$$s.t. \quad |P_i| \leq (1 + \varepsilon)\frac{|V|}{k}, \forall i \qquad (2)$$

$$|Q_i| \leq (1 + \eta), \forall i \qquad (3)$$

The objective function minimizes the number of communication links between two different subgraphs or protection-zones. The damage propagates through the communication links between CPS components, therefore minimizing the communication across protection-zones reduces the chances of inter protection-zone damage propagation. Moreover, the Constraint (2) balances the sizes of the partitions as small sized partitions are unable to restrict the damage spread in one protection-zone and large sized partitions allow damage to propagate to a large number CPS components without any hindrance. Constraint (3) ensures that a CPS component in a protection-zone is at most connected to components in $1 + \eta$ protection-zones. The intuition is that, if a component that is connected to components in many different protection-zones is attacked then damage could quickly propagate to other protection-zones. The parameters $\varepsilon$ and $\eta$ are used to relax the constraints, the lower values of these parameters the better expected performance in terms damage containment, but usually if the constraints are too tight then there might not be a feasible solution for a given CPS topological structure. This graph decomposition problem is similar to the classical uniform graph partitioning problem[8] and the PAM employs a greedy algorithm to create protection-zones (partitions) of similar sizes.

PAM maintains information about the respective member components of protection-zones and the boundary components of all protection-zones. A CPS component that has connections to components in other protection-zones is a boundary component. Figure 2(d) shows the creation of protection-zones within an IB for AMI as an example CPS.

**Intrusion Response System**
IRES is activated by the IDS upon detection of an intrusion. IRES generates response for the cyber-attack by utilizing the information provided by the IDS, PAM and operational logs of CPS. The functionality of IRES is described as follows.

**Damage Assessment (DA).** The DA module uses the protection-zone information provided by PAM to identify the protection-zones where damage has been spread. PAM shares two data structures with the DA module. The first one maps every CPS component to a protection-zone and is named pz_component_list. The second data structure contains information about the boundary components of the protection-zones and is named pz_boundary_set. The IDS detects intrusions in the CPS and generates alerts that specifies the CPS components that are compromised. The following pseudocode describes the damage assessment process.

```
Input: pz_component_list, pz_boundary_set, ids_alert
Output: response_set
1. for each ids_alert of a corrupted CPS component
2. use pz_component_list to find the protection-zone P to which the
   corrupted component belongs
3. identify the boundary components of P using pz_boundary_set
4. include the identified boundary components in response_set
```

For each alert received by the DA module, response_set is computed. The components that need to be isolated from the rest of CPS are in the response_set. It is possible that by the time IDS detects an intrusion the damage has already spread to other protection-zones. In that case, the IDS keeps on generating alerts as more CPS components get compromised and the DA module keeps generating response_sets and provide them to the response manager module until the damage is contained.

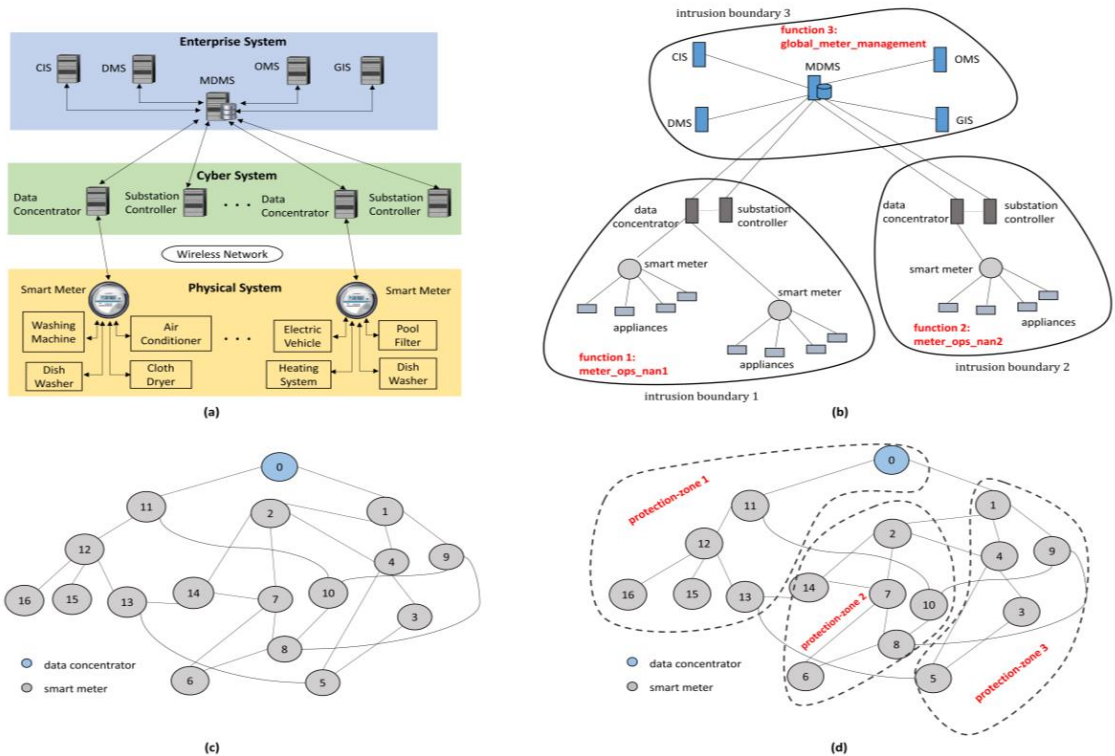

Figure 2. Partitioning process for AMI. (a) Multi-layered system of systems AMI architecture. (b) Demarcation of IBs on a given AMI topology. (c) AMI topological structure in an IB. (d) Partitioning of the AMI topological structure to form protection-zones.

**Response Manager.** Response manager coordinates the operations of DA and interfaces with the CPS control system. Response manager starts its operations the moment it receives an alert from IDS. First, the response manager halts the operation of the compromised components. Subsequently, it sends instructions to CPS control system to disconnect the CPS components identified in response_set. The goal here is to stop inter protection-zone damage propagation. This action might temporarily halt certain CPS functions, resulting in short-lived unavailability of those functions in exchange for the benefit of damage containment. Response manager continues this process until there are no more alerts from the IDS, indicating the damage has been successfully contained. Subsequently, the response manager activates IREC to start the recovery process. Additionally, response manager sends IRES operational information to PEM.

**Intrusion Recovery System**
IREC recovers the system in an automated fashion by determining the exact extent of damage, and performing corrective actions accordingly. The architecture responds first and recovers afterwards, because the recovery is a time consuming process while response can be deployed quickly after the detection of the cyber-attack. We explain the functions of several IREC modules briefly.

**Recovery Analysis (RA).** RA module assesses the extent of damage by precisely identifying the compromised components of CPS. For an IDS alert, the recovery process is initiated after completion of the response process. RA module analyzes operational logs of the CPS that contains detailed information about the activities of CPS components. The activity of interest is usually the data flow information between CPS components. RA module uses the IDS alert information to find the start time of the cyber-attack ($T_s$) and the compromised CPS component. RA module then starts analyzing the CPS operational log to identify all of the components to which the compromised component communicated and includes them in a data structure known as repair_set. The RA process is described in the following pseudocode.

```
Input: ids_alert, op_log
Output: repair_set
1. for each ids_alert of corrupted CPS component
2.   include compromised component in repair_set
3. for each new entry s in repair_set
4.   identify components r in op_log that communicated with  s after T_s
5.   include  component  c  ∈  r  in  repair_set, if  c  ∉  repair_set
```

It can be observed in the pseudocode that the RA module generates the repair_set by building causality relationship between components using inter-component communication information.

**Recovery Protocol Knowledgebase (RPK).** Once the RA module identifies the compromised components, the RPK module determines the recovery protocol to recover them. RPK module maintains a knowledgebase that provides procedures that needs to be followed to recover the compromised CPS components. For example, if the data management server is compromised then the recovery techniques mentioned in[9] can be employed. Similarly, if a field device is compromised then a redundant device could be activated, if available. Otherwise, if the device was providing essential data to a controller then instructions can be given to the controller to use simulated data till the device is restored manually.

**Recovery Manager.** Similar to response manager, the recovery manager interfaces with CPS control system. Recovery manager directs recovery instructions to the CPS controllers to take corrective actions based on the information provided by the RPK module, for a particular attack. Notice that recovery manager might not be able to completely recover the CPS functionality in an automated manner due limitations of the CPS infrastructure or unavailability of appropriate recovery protocol in RPK for a particular component. In that case, the recovery manager only disables the corrupted component, and generates an alert to recover the system manually. Once the recovery process of the system is complete, the normal communication of all the CPS components is restored. Recovery manager also records the activities of IREC and forwards this information to PEM.

**Performance Monitor**
PEM provides an overview of response and recovery system performance by collecting statistics from IRES and IREC, and calculating different performance metrics. The performance monitor has two modules that we discuss briefly.

**Statistics Collection.** Information from different systems of the proposed security architecture is collected and analyzed by statistics collection module. The received information can be in the form of extensive log files, therefore this module extracts useful data that is used by metrics calculation module.

**Metrics Calculation.** We introduce ω as a measure of serviceable CPS functions, and derive δ from ω that reflects the impairment of CPS functions caused by an attack. The metrics are defined as follows.

- Operational availability (ω): the available CPS functionality in terms of percentage of components unaffected in an attack for a CPS function.
- Damage extent (δ): the compromised CPS functionality in terms of percentage of components damaged in an attack for a CPS function.

Note that these metrics concern a particular CPS function. The overall CPS availability and damage extent can be found out by averaging the metric for all CPS functions. Also note that we can compute one metric from the other by subtracting the known metric value from 100.

## Simulations and Evaluation

We have used AMI as the target CPS application to evaluate the performance of the proposed security architecture. We first discuss the threat model concerning AMI followed by a discussion on the simulation tool and implementation of proposed architecture. Finally, we present our experimental setup and compare the performance of the proposed architecture with the commonly adapted intrusion prevention approach.

**Threat Model and Implementation**
We have employed the electricity pricing manipulation cyber-attacks on AMI[10] as the attack scenario in our experiments. The attacker tries to exploit access vulnerabilities of smart meters through a direct physical link or through the AMI. The objective of the attacker in electricity pricing manipulation cyber-attack is to disrupt the distribution system of the power grid such that electricity supply to the targeted regions is suspended. The attacker maliciously gains access to a smart meter and usually injects a malware that could be propagated through the network. In AMI, smart meters are connected to the data concentrator through a mesh network that forms a neighborhood area network (NAN). The target of the malware is the guideline price information that is used by the smart controller to manage the usage of different appliances to reduce the electricity bill. The smart controller when observes a lower price at a certain time of the day, schedules the usage of many appliances, resulting in the increase in energy consumption of that particular consumer. The consumers in a NAN are usually connected to the same bus used by the distribution system of the power grid. If the attacker succeeds to manipulate the pricing information in a large number of smart meters one after the other, the transmission line that delivers power to the bus can trip. This can cause the transmission system to reallocate the power distribution to other transmission lines connecting the bus, which in turn can trip due to excessive load. This results in a cascading failure and the community connected to the bus can be out of power. This disruption in the distribution system generally doesn't stay within a community as from the generation system to the consumers, power flows through transmission lines linked to several buses as a result a total blackout can occur.

In order to simulate this scenario we have used SecAMI[11], an opensource simulator developed to study the impact of cyber-attacks on AMI. SecAMI can be used to perform two operations: first an AMI topology involving smart meters and data concentrators can be created as an undirected graph, second an attack on the created topology can be simulated that compromises one node (smart meter) after the other according to three simulation parameters – compromise time of a node (ct), hop time from one node to another (ht), and detection time of a compromised node (dt). These parameters determine the attack propagation speed with respect to the detection speed in the simulation. We have modified the SecAMI to implement the protection-zone aware response mechanism of the architecture. To create protection-zones (partitions) for a given AMI topology (undirected graph), we have built a partitioning module on top of SecAMI that takes a parameter k and a topology graph as inputs to create k protection-zones of equal sizes. The partitioning module maintains information about the members of a protection-zone and boundary components of the created protection-zones. SecAMI uses a general response mechanism in which as a node is detected to be compromised, it is disconnected from the rest of the network. The proposed security architecture's response mechanism uses the protection-zone information to identify the protection-zones whose member nodes are detected to be compromised. Subsequently, it immediately halts the communication of the boundary nodes of the under attack protection-zones from the rest of the AMI topology.

**Experimentation and Results**
To evaluate the performance of the proposed security architecture in terms of damage containment, we have conducted several experiments. The variable parameters in the experiments are: the size of the topology in terms of number of nodes (n), the compromise to detection time ratio ($\sigma$ = ct/dt) that specifies the attack propagation speed with respect to detection speed, and the partitioning parameter (k) that specifies the number of protection-zones for a given topology. The performance of the proposed architecture is measured in terms of three metrics. The first two are aforementioned general CPS damage containment evaluation metrics i.e. $\omega$ and $\delta$. The third metric is AEL that is specific to AMI and power grid analysis. In our experiments, we consider attacks on function 1: meter_ops_nan1 i.e. in intrusion boundary 1 in Figure 2(b). Since only one function is considered, $\omega$ is the percentage of components unaffected in the attack and $\delta$ is the percentage of components compromised in the attack. For AEL we assume that each smart meter in the topology is connected to a medium-sized household with standard appliances like washing machine, dish washer etc. The smart controller of each consumer schedules the energy consumption for next 24h based on the guideline pricing information. The quadratic pricing model presented in[12] is employed to form the normal guideline price and energy consumption for a general household.

The value of $\omega$ for different values of n with $\sigma$ = 3 and k = 0, 8 is shown in Figure 3(a).

It can be seen that the performance of the proposed partition-driven architecture (k=8) is considerably higher as compared to the non-partitioned system (k=0) e.g. at n=200, the value of ω is 84.5 for the partitioned system while its value is 7.4 for the non-partitioned system. Moreover, note that the value of ω increases as the value of n increases until 200 nodes after that it decreases. This is due to the fact that there exists an optimal protection-zone size that maximizes the benefit. If the size of the protection-zone is too small, it is more likely that damage will propagate to other protection-zones. If the size is too big then many member nodes of the protection-zone can get corrupted without any hindrance. Figure 3(b) shows the values of δ for 200 node topology for different σ and k values. Notice that even when the attack propagation speed is high i.e. lower values of σ, the partitioned system (k>0) can effectively contain damage relative to non-partitioned system (k=0). For higher σ values, there is ample time for the response mechanism to react and hence the performance of the partition-driven architecture is significantly better. Figure 3(c) shows the manipulated guideline price values in a smart meter[10]. This change in guideline price increases the energy consumption of a consumer by 1.3 kWh on average in our simulations. Figure 3(d) shows the increase in overall demand by all the customers corresponding to 200 node topology for non-partitioned (k=0) and partitioned systems (k=8). It can be noted that the partition-driven architecture efficiently handles the attack and correspondingly the load on the transmission lines connected to the bus connecting 200 consumers is not drastically increased in comparison to the non-partitioned system.

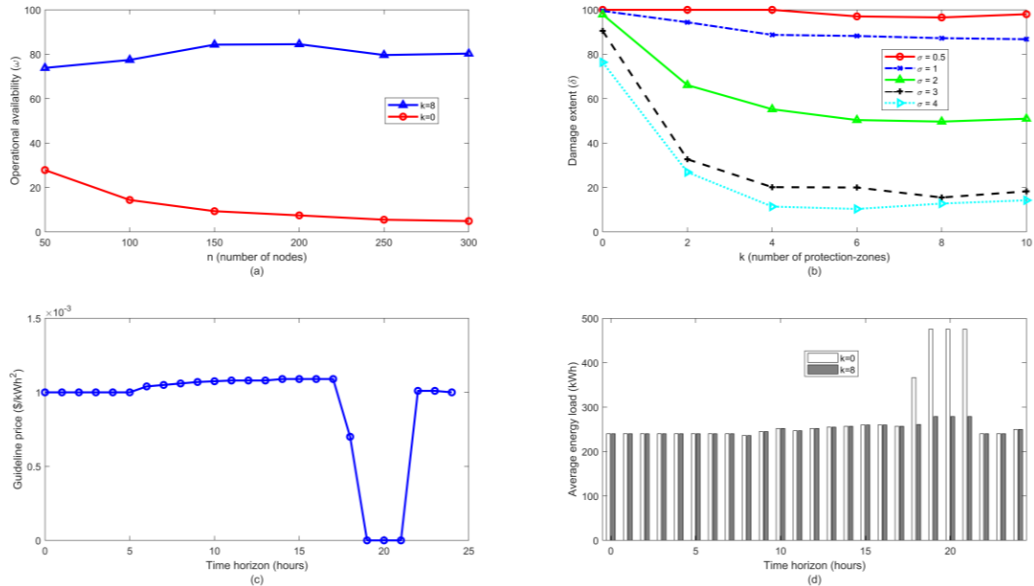

Figure 3. Proposed partition-driven security architecture performance. (a) Operational availability over topology size for σ = 3. (b) Damage extent over number of protection-zones for n = 200. (c) Guideline electricity price in pricing manipulation cyber-attack. (d) Overall AEL consumption of a community for n = 200, σ = 3.

The partition-driven security architecture presented in this article proposes a comprehensive security solution for CPSs. The proposed architecture is able to contain the damage propagation and recover the CPS automatically. However, there are several challenges that must be addressed to develop a robust security solution for CPSs. These include dealing with the performance of IDS and evaluation of the proposed security architecture for more complex attacks.

**References**

1. I. Eusgeld, C. Nan, and S. Dietz, "Sytem-of-Systems Approach for Interdependent Critical Infrastructures," *Reliability Engineering and System Safety*, vol. 96, no. 6, 2011, pp. 679-686.

2. "Framwork for Cyber-physcial Systems Release 1.0", NIST Cyber-physical systems Public Working Group, May 2016; pages.nist.gov/cpspwg.

3. M. Biro, A. Mashkoor, J. Sametinger, and R. Seker, "Software Safety and Security Risk Mitigation in Cyber-physical Systems," *IEEE Software*, vol. 35, no. 1, 2017, pp. 24-29.

4. A. Humayed, J. Lin, F. Li, and B. Luo, "Cyber-Physical Systems Security—A Survey," *IEEE Internet of Things Journal*, vol. 4, no. 6, 2017, pp. 1802-1831.

5. S. Zonouz, H. Khurana, W. Sanders, and T. Yardley, "RRE: A Game-Theoretic Intrusion Response and Recovery Engine," *IEEE Transactions on Parallel and Distributed Systems*, vol. 25, no. 2, 2014, pp. 395-406.

6. R. Mitchell, and I. Chen, "A Survey of Intrusion Detection Techniques for Cyber-Physical Systems," *ACM Computing Surveys*, vol. 46, no. 55, 2014.

7. "Snort – network intrusion detection system,"; https:// www.snort.org.

8. B. Kernighan, and S. Lin, "An Efficient Heuristic Procedure for Partitioning Graphs," *The Bell System Technical Journal*, vol. 49, no. 2, 1970, pp. 291-307.

9. P. Liu, "Architectures for intrusion tolerant database systems," Proc. 18$^{th}$ Annual Computer Security Applications Conference, 2002.

10. Y. Liu, S. Hu, and A. Zomaya, "The Hierarchical Smart Home Cyberattack Detection Considering Power Overloading and Frequency Disturbance," *IEEE Transactions on Industrial Informatics,* vol. 12, no. 5, 2016, pp. 1973-1983.

11. T. Shawly, J. Liu, N. Burow, and S. Bagchi, "A Risk Assessment Tool for Advanced Metering Infrastructures," Proc. 2014 IEEE Conference on Smart Grid Communications, 2014.

12. L. Liu, Y. Liu, L. Wang, A. Zamaya, and S. Hu, "Economical and Balanced Energy Usage in the Smart Home Infrastructure: A Tutorial and New Results," *IEEE Transactions on Emerging Topics in Computing*, vol. 3, no. 4, 2015, pp. 556-570.